\documentclass[letterpaper,12pt]{article}
\usepackage{calc,layout,epsfig,setspace,palatcm,booktabs,rotating,multirow,hyperref}
\title{The $E1$ \& $M1$ Spontaneous Decay Rates for an Emitter Inside a Cavity Within a Medium}
\author{Jaideep Singh\footnote{singhj AT nscl DOT msu DOT edu}\\ National Superconducting Cyclotron Laboratory\\ Michigan State University\\Version 1.00}
\date{\today}
\normalsize
\newcommand{\ket}[1]{\ensuremath{\left | #1 \right > }}
\newcommand{\bra}[1]{\ensuremath{\left < #1 \right | }}

\begin{document}
\maketitle
\begin{abstract}
This short note is based on an early draft of a paper we wrote describing our measurement of the hyperfine quenching rate of the clock transition in Yb-171, see \bf \href{http://dx.doi.org/10.1103/PhysRevLett.113.033003}{Phys. Rev. Lett. 113 033003 (2014)} \rm \cite{hfq}.
We discuss the $E1$ and $M1$ spontaneous decay rates of the an emitter residing inside of a real cavity carved out of a vast, uniform, homogenous, isotropic, linear, lossless, dispersionless, and continuous medium.
The ratio of the medium rate to vacuum rate is given by $\Gamma_m/\Gamma_0 =  [G(u)]^2  n^3 / u$, 
where $G(u) = 3u/(2u+1)$ is the local field correction factor, $n = \sqrt{\epsilon\mu/(\epsilon_0\mu_0)}$ is the index of refraction of the medium, $\epsilon(\epsilon_0)$ is the electric permitivity of the medium (vacuum),
$\mu(\mu_0)$ is the magnetic permeability of the medium (vacuum), and $u = \epsilon/\epsilon_0$ for $E1$ transitions or $u = \mu_0/\mu$ for $M1$ transitions.
\end{abstract}
\tableofcontents
\section{Introduction}
Spontaneous emission occurs when an emitter, such as an excited atom or molecule, couples to the zero-point energy fluctuations of the electromagnetic field and undergoes a quantum transition resulting in a photon.
The rate of these transitions, or equivalently the lifetime of the excited state,  depends not only on the intrinsic properties of the emitter, but also on the properties of the surrounding environment.
Since Purcell first pointed this out in 1946 \cite{purcell46}, there have been many experimental demonstrations of this so-called ``Purcell effect'' \cite{drexhage70,yablonovitch88,harklep89} 
and it is one of the hallmarks of cavity quantum electrodynamics.
In recent years, materials are being engineered with a microscopic structure designed specifically to enhance the spontaneous emission rate in order to realize, among other things, miniature lasers and single photon sources \cite{yablonovitch87,lodahl04,inam11}.
Even with all this prodigious effort, there is still considerable tension in the literature about how the spontaneous emission rate of an emitter is modified by the medium that surrounds it \cite{toptygin03}.

\section{Previous Work on Dielectric Media}
This problem has been most extensively studied for electric dipole ($E1$) transitions in homogenous, uniform, isotropic, lossless, and linear dielectric media with dielectric constant $\epsilon$ and index of refraction $n = \sqrt{\epsilon/\epsilon_0}$.
The main difficulties in addressing this question are understanding the way in which the emitter modifies the electromagnetic field in the surrounding medium and the way in which the medium modifies the properties of the emitter.
The former being the main challenge in choosing the appropriate theoretical framework within which to calculate the spontaneous emission rate and the latter being the main challenge in interpreting experiments.

The two main types of theoretical approaches are categorized as \emph{macroscopic} and \emph{microscopic}.
The main difference between these two approaches has to do with whether the medium is treated as continuous (macroscopic) or discrete (microscopic).
In macroscopic theories, first, the electromagnetic field is quantized in the usual way \cite{na76,gl91,bhl92,bhlm96}, but now accounting for the proper normalization of the electromagnetic field fluctuations due to the permittivity of the medium.
Then, to calculate the spontaneous decay rate, one applies Fermi's golden rule noting that the photon density of states scales as the wave number cubed $k^3$, which implies an $n^3$ scaling.
Putting this altogether, accounting for the medium modification of the emitter fluorescence spectrum \cite{stoneham}, and introducing the local field correction \cite{fox12} gives the ratio of the spontaneous decay rate in medium ($q_m \Gamma_m$) to 
that in vacuum ($\Gamma_0$)
\begin{equation}
\frac{q_m \Gamma_m}{\Gamma_0} =  \left ( \frac{\omega_m}{\omega_0} \right)^3 \left ( \frac{D_m}{D_0} \right)^2 n \left [  G(n^2) \right ]^2
\end{equation}
where $\omega$ is the frequency of the transition, 
$D$ is  dipole matrix element, 
and the subscript $m$ ($0$) refers to the medium (vacuum).
The factor $q_m$, known as the quantum efficiency of the transition, is the ratio of the spontaneous emission rate to the total excited state decay rate in medium.
It is included to account for the possible presence of nonradiative decay mechanisms induced by the medium.

The factor $G(u)$, known as the local field correction, is the ratio of the local field immediately surrounding the emitter and the macroscopic field in the dielectric medium far from the emitter.
It is usually calculated by assuming the emitter is a point dipole that is located at the center of a dielectric sphere surrounded by the medium.
Depending on dielectric constant of this sphere, $G(u)$ is either $G_\mathrm{RCM} = 3u/(2u+1)$ assuming the \emph{real cavity model} (RCM) or $G_\mathrm{VCM} = (u+2)/3$ assuming the \emph{virtual cavity model} (VCM).
The choice of model is determined by how similar the emitter is to the medium, or, alternatively, by whether the emitter is located at a substitutional site (RCM) or an interstitial site (VCM) within a discrete lattice \cite{deVries98}. 

In microscopic theories, the medium atoms and emitter atoms are considered separately and couple only via the electromagnetic field.
These calculations typically rely on quantum many body calculations which attempt to account for correlations between nearby medium atoms induced by the presence of the emitter atom \cite{f99,cb00,bm04,fb05,krg11}.
The modification of the spontaneous decay rate in these theories is usually written in one of the two following ways
\begin{equation}
\frac{q_m \Gamma_m}{\Gamma_0} = \left ( \frac{\omega_m}{\omega_0} \right)^3 \left ( \frac{D_m}{D_0} \right)^2 n^k G(n^2)
 = \left ( \frac{\omega_m}{\omega_0} \right)^3 \left ( \frac{D_m}{D_0} \right)^2 \left [ 1 + c_1 \alpha_0 N + c_2 (\alpha_0 N)^2 + \mathcal{O}(\alpha_0 N)^3\right ] 
\end{equation}
where $k=0\ \mathrm{or}\ 1$, 
now $G$ is linear and not quadratic,
$\alpha_0$ is the electric polarizability of the medium atoms, 
$N$ is the number density of the medium atoms, 
and $c_1$ \& $c_2$ depend on the details of the calculation.
We note that the units of $\alpha_0$ are such that it satisfies the Clausius-Mossoti relation given by $\alpha_0 N = 3(n^2-1)/(n^2+2)$.
In all cases, a larger index of refraction is expected to result in a faster rate.

The index of refraction dependence of the spontaneous decay rate has been measured in rare earth ion-complexes in liquids \cite{rikken95},
dye molecules dissolved in water droplets suspended in hydrophobic liquids \cite{lavallard96,lamouche99}, 
rare earth ion-complexes in dense gases \cite{schuurmans98}, 
nanoparticles in air \& liquids \cite{meltzer99}, 
rare earth ions in binary glasses \cite{kumar03, kumar09, duan11}, 
quantum dots in liquids \cite{wuister04}, 
and rare earth ions in vacuum and crystalline solids \cite{duan06, zych12}.
Challenges in interpreting these results include determining the index of refraction of the composite systems, 
the quantum efficiency of the emitter, 
and the changes in the dipole matrix element of the emitter due to the surrounding environment.
In short, there is not a strong consensus in either the theory or experimental community about the correct solution to this problem.
The current status of this problem has been recently reviewed by Dolgaleva \& Boyd \cite{db12}.

\section{Simplifying Assumptions}
In order to make an unambiguous calculation, we restrict ourselves to $E1$ and $M1$ spontaneous decay rates of the an emitter residing inside of a cavity carved out of a vast, uniform, homogenous, isotropic, linear, lossless, dispersionless, and continuous medium.
We assume that the emitter density is sufficiently dilute that we do not need to worry about the effect of a neighbhoring emitter.
A ``vast'' medium simply means that the emitter is far from the boundaries of the medium.
The constraint that the medium is uniform, homogenous, isotropic, linear, lossless, and dispersionless means that the electromagnetic properties of the medium can be described by two scalar, real, frequency independant constants: 
the electric permittibity $\epsilon$ and the magnetic permeability $\mu$.
The index of refraction of the medium is therefore $n = \sqrt{\epsilon \mu/(\epsilon_0 \mu_0)}$, where the subscript $0$ labels the vacuum values.
By placing the emitter inside of a cavity that is carved out of a continuous medium, we are able to unambiguously determine the local field correction factor, which is given by the real cavity model as $G(u) = 3u/(2u+1)$.
In addition, we will simply assume that the oscillator strength and frequency of the transitions are not modified by the medium.

\section{Atomic Electric Dipole Transitions}
We start by Fermi's Golden Rule which gives the tranistion rate due to an operator $\mathcal{H}$ averaged over initial states and summed over final states:
\begin{equation}
	\Gamma = \frac{2\pi}{\hbar} \frac{\sum_a^{N_a}}{N_a} \sum_b^{N_b} \left | \left < b \left | \mathcal{H} \right | a \right > \right |^2 \rho_{a \rightarrow b}
\end{equation}
where $\hbar$ is the Planck constant divided by $2 \pi$, $a(b)$ labels an initial (final) state, $N_{a(b)}$ is the total number of initial (final) states, and $\rho$ is density of available states (number of states per unity energy).
The transition matrix element for electromagnetic transitions involves the dot product between a dipole moment operator and an electromagnetic field.
The dipole momentum operator acts only on the wave function representing the emitter, while the matrix element involving the electromagnetic field is related to the RMS fluctuations of the zero-photon states of the electromagnetic field.

We'll first consider $E1$ transitons for atomic emitters where the electric dipole moment is $\vec{d}$ and the electric field is $\vec{E}$:
\begin{eqnarray}
	\left | \left < b \left | \mathcal{H} \right | a \right > \right |^2 
 & = & \left | \left < b \left | \vec{d}\cdot\vec{E} \right | a \right > \right |^2 
 =  \left | \left < b \left | d_x E_x + d_y E_y + d_z E_z \right | a \right > \right |^2 \\ 
 & = & \left | \left < b \left | d_x \right | a \right > \right |^2 \left | \left < E_x \right > \right |^2
+ \left | \left < b \left | d_y \right | a \right > \right |^2 \left | \left < E_y \right > \right |^2
+ \left | \left < b \left | d_z \right | a \right > \right |^2 \left | \left < E_z \right > \right |^2 \\
 & = & \left | \left < b \left | d_x \right | a \right > \right |^2 \frac{\left | \left < \vec{E} \right > \right |^2}{3}
+ \left | \left < b \left | d_y \right | a \right > \right |^2 \frac{\left | \left < \vec{E} \right > \right |^2}{3}
+ \left | \left < b \left | d_z \right | a \right > \right |^2 \frac{\left | \left < \vec{E} \right > \right |^2}{3} \\
 & = & \frac{1}{3}\left | \left < b \left | \vec{d} \right | a \right > \right |^2 \left | \left < \vec{E} \right > \right |^2
\end{eqnarray}
where we've taken advantage of the orthogonality of electric dipole moment vector operator and the isotropy of space.

In order to determine the matrix element related to the electric field, we first recall that the energy density of a classical electric field is given by:
\begin{equation}
	\frac{\vec{D}\cdot\vec{E}}{2} = \frac{\epsilon}{2} \left | \vec{E} \right |^2
\end{equation}
where $\epsilon$ is the dielectric constant.
This energy density can be related to the zero point energy $\hbar \omega_m/2$ of a simple harmonic oscillator summed over each mode labeled by $m$ \cite{josab}:
\begin{equation}
	\frac{\epsilon}{2} \left | \left < \vec{E} \right > \right |^2 = \frac{1}{2V} \sum_m^{N_m} \frac{\hbar \omega_m}{2}\ \ \ \rightarrow\ \ \ \left | \left < \vec{E} \right > \right |^2 = \sum_m^{N_m} \frac{\hbar \omega_m}{2 \epsilon V}
\end{equation}
where the $2$ in the denominator indicates that the energy is partioned equally between the electric \& magnetic fields, and $V$ is the volume of the space occupied by all the modes.
This sum and volume only make sense when combined with the density of photon states, which will be discussed shortly.
What we've calculated so far involves the fluctuations of the electric field far from the emitter.  
However, as mentioned before, the electric field very close to the emitter are modified by a local field correction factor $G = E_\mathrm{local}/E_\mathrm{far}$.
In our case, it is appropriate to invoke the real cavity model \cite{vv40}:
\begin{equation}
G(u) = G_\mathrm{RCM}(u) = \frac{3u}{2u+1}\ \ \ \ \ \rightarrow\ \ \ \ \ G(\epsilon/\epsilon_0) = \frac{3(\epsilon/\epsilon_0)}{2(\epsilon/\epsilon_0)+1}
\end{equation}
We note that $G(u)$ varies from $1$ to $3/2$ for $u \ge 1$.

We now return to the density of photon states. 
First, the total number of photon states in the volume $V$ with can contain a maximum wavelength $\lambda$ or alternatively minimum wave number $k$ is given by:
\begin{equation}
	2N_m = 2 V \left ( \frac{4 \pi}{3} \right ) \left ( \frac{k}{2\pi} \right )^3
\end{equation}
where the $2$ indicates two polarization modes and we're assuming that $V$ is very large.
The relationship between wave number and frequency is given by:
\begin{equation}
	\nu \lambda = \left ( 2 \pi \nu \right ) \left ( \frac{\lambda}{2 \pi} \right ) = \frac{\omega}{k} = \frac{c}{n} 
\end{equation}
where $c$ is the speed of light in vacuum.
The density of modes near $k$ is given by \cite{loudon}:
\begin{equation}
	\rho_{a \rightarrow b} = \frac{ \partial (2N_m)}{\partial (\hbar \omega)} = \frac{n}{\hbar c} \frac{\partial (2N_m)}{\partial k} = \frac{2Vn}{\hbar c} \left ( \frac{4 \pi}{3} \right ) \frac{3k^2}{(2\pi)^3}
\end{equation}
Finally, we now stipulate that the only modes that contribute to the sum $\sum_m^{N_m}$ are the ones with energy nearly equal to $|E_b - E_a| = \hbar \omega = \hbar k c/n$. 
Putting this altogether gives:
\begin{equation}
	\left | \vec{E}_\mathrm{local} \right |^2 \rho_{a\rightarrow b} = \frac{G^2 n \omega}{\epsilon c} \left ( \frac{4 \pi}{3} \right ) \frac{3k^2}{(2\pi)^3} = \frac{G^2 n^3 \omega^3}{2 \epsilon \pi^2 c^3}
\end{equation}
where the final answer no longer depends on $V$.

Before writing down the $E1$ spontaneous decay rate in medium, it is convenient to recall the following equations for the fine structure constant and the oscialltor strength:
\begin{eqnarray}
	\alpha & = & \frac{e^2}{4 \pi \epsilon_0 \hbar c} \\
	f & = & \frac{2m\omega}{3\hbar} \frac{\sum_a^{N_a}}{N_a} \sum_b^{N_b} \left | \left < b \left | \vec{r} \right | a \right > \right |^2 
\end{eqnarray}
where $e$ is the elementary charge, $m$ is the electron mass, and $\vec{d} = e \vec{r}$.
With these substitutions, we find the $E1$ spontaneous decay rate for an excited atom is:
\begin{eqnarray}
	\Gamma_{E1} & = & 2 \alpha f \left [ \frac{ \left \{ G(\epsilon/\epsilon_0)\right \}^2 n^3}{\epsilon/\epsilon_0} \right ] \left [ \frac{\hbar \omega^2}{m c^2} \right ] \\
& = & \left [ \frac{f}{15.0\ \mathrm{ns}}  \right ] \left [ \frac{ \left \{ G(\epsilon/\epsilon_0)\right \}^2 n^3}{\epsilon/\epsilon_0} \right ] 
\left [ \frac{1000\ \mathrm{nm}}{\lambda} \right ]^2 
\end{eqnarray}
\section{Nuclear Spin Magnetic Dipole Transitions}
We can perform an analgous calculation for $M1$ transitions, which involves the dot product $\vec{\mu}\cdot\vec{B}$, where $\vec{\mu}$ is the magnetic dipole moment and $\vec{B}$ is the magnetic field.
Following the energy density argument from before \cite{josab}, we find:
\begin{equation}
	\frac{\vec{B}\cdot\vec{H}}{2} = \frac{1}{2\mu} \left | \vec{B} \right |^2 = \frac{1}{2V} \sum_m^{N_m} \frac{\hbar \omega_m}{2}\ \ \ \rightarrow\ \ \ \left | \vec{B} \right |^2 = \sum_m^{N_m} \frac{\hbar \omega_m \mu }{2V}
\end{equation}
wher $\mu$ is the magnetic permeability of the medium.
The local field correction factor has the same form with $\epsilon \rightarrow 1/\mu$ \cite{jackson} and we find:
\begin{equation}
G(\mu_0/\mu) = G_\mathrm{RCM}(\mu_0/\mu) = \frac{3}{2+(\mu/\mu_0)}
\end{equation}
where $\mu_0$ is the permeability of the vacuum.
The product of the local $B$-field matrix element squared and the photon density of state (which is the same) is
\begin{equation}
	\left | \vec{B}_\mathrm{local} \right |^2 \rho_{a\rightarrow b} = \frac{G^2 \mu n \omega}{c} \left ( \frac{4 \pi}{3} \right ) \frac{3k^2}{(2\pi)^3} = \frac{G^2 \mu n^3 \omega^3}{2 \pi^2 c^3}
\end{equation}
Putting this altogeher gives:
\begin{equation}
\Gamma_{M1} = \frac{2\pi}{\hbar} \frac{\sum_a^{N_a}}{N_a} \sum_b^{N_b} \left | \left < b \left | \vec{\mu} \right | a \right > \right |^2  \frac{G^2 \mu n^3 \omega^3}{6 \pi^2 c^3}
\end{equation}
where $G$ is now a function of $\mu_0/\mu$ and the factor of $3$ comes from the fact that the fluctuations of the magnetic field are isotropic.

What is remaning now is the calculation of the magnetic dipole matrix element.
For this calculation, we will assume that our emitter is a unpolarized nucleus with spin $S$ and a magnetic moment given by $|\vec{\mu}| = g \mu_N S$, where $\mu_N$ is the nuclear magneton.
The $M1$ transitions that we considering are between the Zeeman levels of the spin that are split by some bias magnetic field giving energy levels $\ \hbar \omega_m = g \mu_N m B$.
We'll start by writing the magnetic dipole vector operator in terms of spin ladder operators:
\begin{equation}
	\vec{\mu} = g \mu_N \vec{S} = g \mu_N \left [ \left ( \frac{S_+ + S_-}{2}\right ) \hat{x} + \left ( \frac{S_+ - S_-}{2i} \right ) \hat{y} + \left ( S_z\right ) \hat{z} \right ] 
\end{equation}
Without loss of generality, we will assume the higher energy state is $\ket{m-1}$, while the lower energy state is $\ket{m}$, which implies that $g>0$.
By noting that $\bra{m-1}S_+\ket{m}=\bra{m-1}S_z\ket{m}=0$, the matrix element is then given by:
\begin{equation}
	\bra{m-1}\vec{\mu}\ket{m} =  g \mu_N \left ( \frac{\bra{m-1}S_-\ket{m}}{2}\right ) \left [ \hat{x} + i \hat{y} \right ] 
\end{equation}
By noting that $\bra{m-1}S_-\ket{m} = \sqrt{S(S+1)-m(m-1)}$, $\sum_{m=0}^{+S} m^2 = S(S+1)(2S+1)/6$, and $\sum_{m=-S}^{+S} m =0$, we find that the amplitude squared averaged over all initial values is:
\begin{equation}
	\frac{1}{2S+1} \sum_{m=-S}^{+S} \left | \bra{m-1}\vec{\mu}\ket{m} \right |^2=  \frac{g^2 \mu_N^2}{2(2S+1)} \sum_{m=-S}^{+S} \left [ S(S+1)-m(m-1) \right ] = \frac{g^2\mu_N^2S(S+1)}{3}
\end{equation}
Putting this altoghether gives the $M1$ spontaneous decay rate for an unpolarized nuclear spin $S$ is a magnetic field $B$:
\begin{eqnarray}
	\Gamma_{M1} & = & \frac{g^2S(S+1)}{9\pi}\left [ \frac{\left \{ G(\mu_0/\mu)\right \}^2 n^3}{\mu_0/\mu} \right ] \left [ \frac{\mu_N^2\mu_0\omega^3}{\hbar c^3} \right ] \\
	& = & \frac{g^5S(S+1)}{9\pi}\left [ \frac{\left \{ G(\mu_0/\mu)\right \}^2 n^3}{\mu_0/\mu} \right ] \left [ \frac{\mu_N^5\mu_0B^3}{\hbar^4 c^3} \right ] \\
	& = & \frac{g^5S(S+1)}{2.28\!\times\!10^{28}\ \mathrm{sec}}\left [ \frac{\left \{ G(\mu_0/\mu)\right \}^2 n^3}{\mu_0/\mu} \right ] \left [ \frac{B}{1\ \mathrm{T}} \right ]^3
\end{eqnarray}
For a sense of scale, using the $g$-factor of an electron in nuclear magneton units $|g_e| = 3681$, $B=10\ \mathrm{T}$, and $\mu=\mu_0$, we find $\Gamma_{M1}^{-1} = 1.43\ \mathrm{yrs}$.
If we were to put in $g \approx 1$, then the resulting decay lifetime would be several orders of magnitude longer than the age of the Universe.
\section{Conclusion}
Determining the spontaneous decay rate in medium is complicated.  The medium can:
\begin{enumerate}
	\item shift the emitter's transition frequency
	\item change the emitter's transition matrix element
	\item introduce non-radiative quenching mechanisms
\end{enumerate}
The first two directly alter the spontaneous decay rate.
The third makes the interpretation of experimental data more challenging.
Furthermore, the medium alters the effect of the fluctuations of the zero-photon state of the electromagnetic field by:
\begin{enumerate}
	\item increasing the density of photon states by a factor of $n^3$
	\item altering the magnitude of the field fluctuations far from the emitter by $\epsilon^{-1}$ or $\mu$
	\item amplifying the magnitude of the field flucuations near the emitter by a local field correction factor $G(u)$ where $u = \epsilon/\epsilon_0$ or $u=\mu_0/\mu$
\end{enumerate}
The first one is well known. 
The last two is a matter of considerable debate in the literature.
The situation becomes even more complicated if the medium is:
\begin{enumerate}
	\item small compared to the transition wavelength: edge effects due boundary conditions 
	\item nonuniform: $\epsilon$ and $\mu$ depend on position
	\item inhomogenous: $\epsilon$ and $\mu$ have to be averaged over different medium species
	\item anisotropic: $\epsilon$ and $\mu$ depend on the direction of emmission
	\item nonlinear
	\item lossy: $\epsilon$ and $\mu$ are complex which implies the medium can reabsorb the emitted light
	\item dispersive: $\epsilon$ and $\mu$ depend on the transition frequency
	\item discrete: some ambiguity about how to quantize the electromagnetic field
\end{enumerate}
If the emitter can be thought of residing in a real cavity carved out of a simple continuous medium and the emitter is unaffected by the medium, 
then the ratio of the in-medium decay rate to the vacuum decay rate is:
\begin{equation}
	\left [ \frac{\Gamma_m}{\Gamma_0} \right ]_\mathrm{RCM} = \left [ \frac{3u}{2u+1}  \right ]^2 \frac{n^3}{u}
\end{equation}
where $n=\sqrt{\epsilon\mu/(\epsilon_0\mu_0)}$ is the index of refraction of the medium and $u = \epsilon/\epsilon_0$ for $E1$ transitions or $u=\mu_0/\mu$ for $M1$ transitions.
For the special case when $\mu = \mu_0$, then $n = \sqrt{\epsilon/\epsilon_0}$, $u = n^2$ for $E1$ transitions, and $u = 1$ for $M1$ transitions, and medium to vacuum decay rate ratios are now:
\begin{equation}
	\left [ \frac{\Gamma_m}{\Gamma_0} \right ]^{E1}_\mathrm{RCM} = \left [ \frac{3n^2}{2n^2+1}  \right ]^2 n\ \ \ \ \ \& \ \ \ \ \ 
	\left [ \frac{\Gamma_m}{\Gamma_0} \right ]^{M1}_\mathrm{RCM} = n^3
\end{equation}
which agrees with \cite{na76,rikken95} but not \cite{gl91}.
Based on the energy density scaling argument used in \cite{josab}, we believe our solution is correct.
\bibliography{m1}
\bibliographystyle{unsrt}
\end{document}